\pdfoutput=1
\documentclass[reprint,12pt,tightenlines,amsmath,amssymb,aip,jmp,nofootinbib,longbibliography,showkeys]{revtex4-2}

\usepackage{amsthm}
\usepackage{hyperref}
\hypersetup{colorlinks,allcolors=blue,bookmarksopen=true}
\usepackage[T1]{fontenc}

\usepackage{macros/my_macros}
\usepackage{macros/my_formatting}
\usepackage{macros/popupcite}

\begin{document}

\title{Physical quantities as a partially additive field}

\author{Georgy Alymov}
\email{alymov@phystech.su}
\affiliation{Moscow Institute of Physics and Technology (National Research University), Dolgoprudny 141700, Russia}

\begin{abstract}
  We generalize the concept of a field by allowing addition to be a partial operation. We show that elements of such a ``partially additive field'' share many similarities with physical quantities. In particular, they form subsets of mutually summable elements (similar to physical dimensions), dimensionless elements (those summable with 1) form a field, and every element can be uniquely represented as a product of a dimensionless element and any non-zero element of the same dimension (a unit). We also discuss the conditions for the existence of a coherent unit system. In contrast to previous works, our axiomatization encompasses quantities, values, units, and dimensions in a single algebraic structure, illustrating that partial operations may provide a more elegant description of the physical world.
\end{abstract}


\maketitle
\section{Introduction}\label{sec:intro}

In the first decades after the notion of a physical quantity was introduced by
Maxwell {\cite{Maxwell}}, it was used more or less intuitively: every
scientist knew how to work with quantities, but the axioms of quantity
calculus were not stated explicitly {\cite{Bridgman}}. Many scientists
believed that calculations can be performed only with the numerical values of
quantities in some unit system, not with the quantities themselves
{\cite{deBoer1995review}}.

Roughly since the 1950s--1960s, numerous works started to emerge that attempt
to put quantity calculus on a rigorous axiomatic basis
{\cite{deBoer1995review,Fleischmann1951,Drobot1953,Quade1961,Parkinson1964,Kurth1965,Whitney1968,Steiner1969,Krantz1971,Quade1971,Bunge1971,Page1975,Brunk1975,Szekeres1978,Carlson1979,Rychlewski1980,Houard1981,Carinena1988,Hart1995}}.
This area of research remains active to this day
{\cite{Aragon2004,Pescetti2009,Sharlow2009,Janyska2010,Tao2012,Kitano2013,Domotor2017,Raposo2018,ZapataCarratala2022,Jonsson2014,Jonsson2023}}.

Most of these axiomatizations either use the existence of physical dimensions
and/or units as a postulate,\footnote{Disguised under various names: ``kinds of
quantities'' {\cite{deBoer1995review}}, ``Gr{\"o}{\ss}enarten''
{\cite{Fleischmann1951}}, ``Vektorr{\"a}ume von V-Elementen''
{\cite{Quade1961}}, ``equivalence classes'' {\cite{Steiner1969}}, ``basic
dimensions values'' {\cite{Bunge1971}}, ``coherent units'' {\cite{Page1975}},
``pre-units'' {\cite{Carlson1979}}, ``dimensions''
{\cite{Houard1981,Raposo2018,ZapataCarratala2022}}, ``types''
{\cite{Hart1995}}, ``labels'' {\cite{Aragon2004}}, ``axiation''
{\cite{Sharlow2009}}, ``basic spaces of scales'' {\cite{Janyska2010}},
``one-dimensional real vector spaces'' {\cite{Tao2012}}.} or postulate that
quantities can be multiplied by real (or positive real) numbers and define
dimensions as equivalence classes of quantities that differ by a numerical
factor
{\cite{Drobot1953,Kurth1965,Whitney1968,Krantz1971,Rychlewski1980,Kitano2013,Domotor2017,Jonsson2014,Jonsson2023}}.
Some authors simply define arithmetic operations on quantities via operations
on their numerical values and dimensional exponents
{\cite{Parkinson1964,Pescetti2009}}. An unconventional approach was taken in
{\cite{Szekeres1978,Carinena1988}}, where one of the basic notions is
``gauge'' (essentially the numerical value of a quantity, but defined
axiomatically, without reference to units).

On the other hand, it is known that physical dimensions place restrictions on
which quantities can be added together, and one might start with axiomatizing
addition as a partial binary operation and try to deduce physical dimensions
and units from other axioms, such as commutativity, associativity, and
distributivity. This idea, while obvious, does not seem to have been pursued
in the literature.

In Section \ref{sec:PAFs}, we introduce the concept of a {\tmem{partially additive field}},
which differs from the usual field in four aspects: (1) addition is a partial
operation; (2) axioms such as commutativity are interpreted as ``both sides
are defined and equal, or both sides are undefined''; (3) instead of a single
``global'' zero, each element $a$ has a unique zero $0_a$ such that $a + 0_a =
a$, but different elements may have different zeros; (4) non-triviality axiom
is formulated for each distinct zero separately (in physical terms, ``there
exist non-zero quantities of every physical dimension'').

Then we show that elements of a partially additive field behave very similar
to physical quantities. They form subsets of mutually summable elements
(similar to physical dimensions). Among those, the subset of elements summable
with 1 (dimensionless elements) forms a field. Any element can be uniquely
represented as a product of a ``number'' (a dimensionless element) and a unit
(any nonzero element of the same dimension). Basic arithmetic can be done in
the obvious way provided the unit system forms a group under multiplication.

In Section \ref{sec:quantities}, we introduce the notion of a coherent unit system,
i. e., one that forms a group under multiplication and contains one unit for each
dimension. We discuss the conditions for its existence and propose two
sufficient axioms: (1) no dimensionful roots of dimensionless elements; (2)
the multiplicative group of non-zero dimensionless elements is cotorsion (this
includes the most physically important cases, $\mathbb{R}\backslash \{ 0 \}$
and $\mathbb{C}\backslash \{ 0 \}$). We also discuss other, more ``natural'', but
more restrictive axioms.

The main result of this paper is to show that physical dimensions and physical
units arise naturally from slightly modified axioms of a field and do not have
to be explicitly postulated.

An obvious generalization of {\tmem{partially additive fields}} is to make
multiplication also a partial operation. We name this algebraic structure a
{\tmem{fieldoid}}, by analogy with groupoids, and show in Appendix \ref{sec:fieldoids}
that a {\tmem{fieldoid}} is essentially a union of disjoint partially
additive fields.
\section{Partially additive fields}\label{sec:PAFs}

Here we introduce the concept of a {\tmem{partially additive field}}, which is
similar to a field, but with multiple zeros and addition not always defined.%
\footnote{A similar, but different structure appears in the mathematical
literature under the name of a \emph{partial field}~\cite{partialfields}. A
\emph{partial field} has a global zero and a different axiom of associativity
than our \emph{partially additive field}. Summability is generally not
transitive in partial fields, $\{-1, 0, 1\}$ being a counterexample ($1 + 0 = 1$,
but $1 + 1$ is undefined). Not surprisingly, partial fields have not been
applied to quantity calculus.}
Then we will show how these multiple zeros naturally correspond to the concept
of physical dimensions and that each element of a partially additive field is
a product of a dimensionless element and a unit (any other non-zero element of
the same dimension).

\newpage

It will be convenient to use a special symbol $\mathfrak{u}$ for the results
of undefined operations.

\begin{definition}
  Let $\mathcal{Q}_{\mathfrak{u}}$ be a union of two disjoint sets
  $\mathcal{Q}$ and $\mathcal{U}$, with $\mathcal{U}$ containing a single
  element, $\mathfrak{u}$. Let $+ : \mathcal{Q}_{\mathfrak{u}} \otimes
  \mathcal{Q}_{\mathfrak{u}} \rightarrow \mathcal{Q}_{\mathfrak{u}}$, $\times
  : \mathcal{Q}_{\mathfrak{u}} \otimes \mathcal{Q}_{\mathfrak{u}} \rightarrow
  \mathcal{Q}_{\mathfrak{u}}$ be two binary operations on
  $\mathcal{Q}_{\mathfrak{u}}$.
  
  We will use notation $0_a$ for the zero element corresponding to some $a \in
  \mathcal{Q}$, and $\mathcal{Z}= \left\{ a \in \mathcal{Q}| \exists b \in
  \mathcal{Q} \quad a = 0_b \right\}$ for the set of all zero elements of
  $\mathcal{Q}$.
  
  $\tilde{\mathcal{F}} = (\mathcal{Q}, \mathfrak{u}, +, \times)$ is called a
  partially additive field iff the following axioms are satisfied:
  \begin{eq*}
    \forall a & \in \mathcal{Q}_{\mathfrak{u}} & \mathfrak{u}+ a & = \mathfrak{u}\\
    \forall a & \in \mathcal{Q}_{\mathfrak{u}} & \mathfrak{u} \times a & = \mathfrak{u}\\
    \forall a, b & \in \mathcal{Q} & a \times b & \in \mathcal{Q}\\
    \forall a, b & \in \mathcal{Q}_{\mathfrak{u}} & a + b & = b + a\\
    \forall a, b & \in \mathcal{Q}_{\mathfrak{u}} & a \times b & = b \times a\\
    \forall a, b, c & \in \mathcal{Q}_{\mathfrak{u}} & (a + b) + c & = a + (b + c)\\
    \forall a, b, c & \in \mathcal{Q}_{\mathfrak{u}} & (a \times b) \times c & = a \times (b \times c)\\
    \forall a, b, c & \in \mathcal{Q}_{\mathfrak{u}} & a \times (b + c) & = (a \times b) + (a \times c)\\
    \forall a \in \mathcal{Q} \quad \exists !0_a & \in \mathcal{Q}  & a + 0_a & = a\\
    \exists 1 \in \mathcal{Q} \quad \forall a & \in \mathcal{Q} & a \times 1 & = a\\
    \forall a \in \mathcal{Q} \quad \exists (- a) & \in \mathcal{Q} & a + (-a) & = 0_a\\
    \forall a \in \mathcal{Q}\backslash\mathcal{Z} \quad \exists a^{- 1} & \in \mathcal{Q} & a \times a^{- 1} & = 1\\
    \forall a \in \mathcal{Z} \quad \exists b & \in \mathcal{Q}\backslash\mathcal{Z} & a & = 0_b
  \end{eq*}
\end{definition}

These axioms are similar to the axioms of a field, with the following
differences:
\begin{itemize}
  \item addition may be undefined for some arguments (i. e., may return
  $\mathfrak{u}$);
  
  \item equalities should be read as ``both sides are defined and equal, or
  both sides are undefined'';
  
  \item non-global zeros (unique for every element, but not necessarily
  globally);
  
  \item non-triviality for every distinct zero (i. e., every physical
  dimension).
\end{itemize}
From now on, lowercase Latin letters will denote elements of $\mathcal{Q}$ for
some partially additive field $\tilde{\mathcal{F}} = (\mathcal{Q},
\mathfrak{u}, +, \times)$, unless otherwise stated. We will also use notation
$\tilde{\mathcal{F}}_a = \{ b \in \mathcal{Q}|b + a \neq \mathfrak{u} \}$ for
the set of all elements of $\tilde{\mathcal{F}}$ summable with $a \in
\mathcal{Q}$.

\begin{definition}
  $a$ is called a dimensionless element of a partially additive field
  $\tilde{\mathcal{F}}$ with multiplicative identity $1$ iff $a
  \in \tilde{\mathcal{F}}_1$. All the other elements are called dimensionful.
\end{definition}

\newpage

\begin{lemma}
  $a + b \neq \mathfrak{u} \Leftrightarrow 0_a = 0_b$.\label{identical-zeros}
\end{lemma}

\begin{proof}
  Assume $0_a = 0_b$ and $a + b =\mathfrak{u}$. Then $a = a + 0_a = a + 0_b =
  a + b + (- b) =\mathfrak{u}+ (- b) =\mathfrak{u}$, a contradiction. Therefore,
  $(0_a = 0_b) \Rightarrow (a + b \neq \mathfrak{u}) .$
  
  Assume $a + b \neq \mathfrak{u}$. Then $a + b = a + 0_a + b = a + b + 0_b$.
  Since $a + b$ must have a unique zero, $0_a = 0_b = 0_{a + b}$. 
\end{proof}

\begin{corollary}
  In partially additive fields summability is transitive.
  Moreover, it is an equivalence relation.
\end{corollary}

\begin{lemma}
  $(- a)$ and $a^{- 1}$ are unique.
\end{lemma}

\begin{proof}
  Suppose there are two elements $(- a), (- a)'$ such that $a + (- a) = a + (-
  a)' = 0_a$. They are summable with $a$, so $0_{(- a)} = 0_{(- a)'} = 0_a$.
  Therefore, $(- a) = (- a) + a + (- a)' = (- a)'$. The proof for $a^{- 1}$ is
  similar.
\end{proof}

\begin{lemma}
  $0_a$ is the only zero element in $\tilde{\mathcal{F}}_a .$
\end{lemma}

\begin{proof}
  $0_a \in \tilde{\mathcal{F}}_a$ because it is summable with $a$. If $0_b \in
  \tilde{\mathcal{F}}_a$, then $0_b$ is summable with both $b$ and $a$. By
  transitivity, $a$ is summable with $b$. Therefore, $0_b = 0_a$.
\end{proof}

\begin{lemma}
  Let $a, b$ be two dimensionless elements of a partially additive field
  $\tilde{\mathcal{F}}$. Then $a + b, a \times b, 0_a, 1, - a \in
  \tilde{\mathcal{F}}_1$. If $a \nin \mathcal{Z}$, then $a^{- 1} \in
  \tilde{\mathcal{F}}_1$.
\end{lemma}

\begin{proof}
  $(a + 1) \times (b + 1) = a \times b + a + b + 1$. Since $a + 1 \neq
  \mathfrak{u}$ and $b + 1 \neq \mathfrak{u}$, $a + b$ and $a \times b$ are
  summable with 1.

  $0_a$, $(- a)$, and 1 are summable with $a$ and, by transitivity, with 1.

  $a^{- 1} \times (a + 1) = 1 + a^{- 1} .$ Therefore, $a^{- 1}$ is summable
  with 1.
\end{proof}

\begin{lemma}
  $1 \nin \mathcal{Z}$.\label{non-zero-one}
\end{lemma}

\begin{proof}
  If $1 \in \mathcal{Z}$, then $1 = 0_1$, because $1 \in
  \tilde{\mathcal{F}}_1$ and $0_1$ is the only zero element in
  $\tilde{\mathcal{F}}_1$. Then for any $a \in \mathcal{Q}$ $a = a \times 1 =
  a \times (1 + 0_1) = a \times (1 + 1) = a \times 1 + a \times 1 = a + a$.
  This implies every $a$ is its own zero, which contradicts the non-triviality
  axiom.
\end{proof}

\begin{theorem}
  Dimensionless elements of a partially additive field $\tilde{\mathcal{F}}$
  form a field.\label{dimensionless-field}
\end{theorem}

\begin{proof}
  The previous lemmas establish that $\tilde{\mathcal{F}}_1$ is closed under
  addition, multiplication, and corresponding inversions, and that it contains
  distinct additive and multiplicative identities ($0_1$ and $1$,
  respectively) that are global for all elements of $\tilde{\mathcal{F}}_1$.
  Commutativity, associativity, and distributivity are already in the axioms
  of a partially additive field. Therefore, $\tilde{\mathcal{F}}_1$ is a
  field.
\end{proof}

\newpage

\begin{definition}
  A set containing a single non-zero element from each distinct
  $\tilde{\mathcal{F}}_a$ (i. e., each equivalence class of mutually summable
  elements) is called a unit system. Any element of this set is called a unit.
\end{definition}

\begin{theorem}
  Let $\tilde{\mathcal{F}} = (\mathcal{Q}, \mathfrak{u}, +, \times)$ be a
  partially additive field and $\mathcal{U}$ a unit system in
  $\tilde{\mathcal{F}}$. Then every element $a \in \mathcal{Q}$ can be
  uniquely represented as a product of a dimensionless element $v_a$ and a
  unit $u_a$.\label{value-unit}
  
  In this representation, arithmetic operations obey the following rules:
  \begin{itemize}
    \item $v_a \times u_a + v_b \times u_b = (v_a + v_b) \times u_a$ iff $u_a
    = u_b$, else undefined;
    
    \item $(v_a \times u_a) \times (v_b \times u_b) = (v_a \times v_b) \times
    (u_a \times u_b)$ if $u_a \times u_b \in \mathcal{U}$;
    
    \item $0_{v_a \times u_a} = 0_1 \times u_a$;
    
    \item $1 = 1 \times 1$ if $u_1 = 1 ;$
    
    \item $- (v_a \times u_a) = (- v_a) \times u_a$;
    
    \item $(v_a \times u_a)^{- 1} = v_a^{- 1} \times u_a^{- 1}$ if $u_a^{- 1}
    \in \mathcal{U}$.
  \end{itemize}
\end{theorem}

\begin{proof}
  ({\tmem{Existence}}) Select $u_a \in \mathcal{U}$ such that $u_a \in
  \tilde{\mathcal{F}}_a$. $u_a$ is non-zero by the definition of a unit
  system. Then $a = a \times 1 = a \times u_a^{- 1} \times u_a$. Since $u_a$
  is summable with $a$, $a \times u_a^{- 1} + 1 = (a + u_a) \times u_a^{- 1}
  \neq \mathfrak{u}$. Therefore, $a \times u_a^{- 1}$ is summable with 1,
  i.e., dimensionless.
  
  ({\tmem{Uniqueness}}) If $a = v_a \times u_a$ with $v_a$ dimensionless, then
  $a + u_a = (v_a + 1) \times u_a \neq \mathfrak{u}$. Therefore, $u_a \in
  \tilde{\mathcal{F}}_a$ and is unique by the definition of a unit system.
  $v_a$ is also unique because $v_a = v_a \times 1 = v_a \times u_a \times
  u_a^{- 1} = a \times u_a^{- 1}$.
  
  The addition and multiplication laws immediately follow from (i)
  commutativity, associativity, distributivity, (ii) the subset of
  dimensionless elements being closed under arithmetic operations, and (iii)
  the one-to-one correspondence between units and subsets of mutually summable
  elements.
  
  Because $v_a$ is dimensionless, $0_{v_a} = 0_1$. Therefore, $v_a \times u_a
  + 0_1 \times u_a = (v_a + 0_1) \times u_a = (v_a + 0_{v_a}) \times u_a = v_a
  \times u_a$. Since $0_{v_a \times u_a}$ must be unique, $0_{v_a \times u_a}
  = 0_1 \times u_a$.
  
  $(- v_a) \times u_a + v_a \times u_a = 0_{v_a} \times u_a = 0_1 \times u_a =
  0_{v_a \times u_a}$. Since additive inverse is unique, $- (v_a \times u_a) =
  (- v_a) \times u_a$. Similarly for $(v_a \times u_a)^{- 1}$.
\end{proof}

Now we are ready to define the remaining basic notions of quantity calculus.

\begin{definition}
  Under conditions of Theorem \ref{value-unit}, $v_a$ is called the numerical
  value of $a$ (in the chosen unit system), and $u_a$ its unit. We may sometimes
  omit the adjective ``numerical'' for brevity.
\end{definition}

\begin{definition}
  Dimensions in a partially additive field are equivalence classes of mutually
  summable elements. The dimension of a specific element is the equivalence class
  to which it belongs.
\end{definition}

Theorem \ref{value-unit} implies that mutually summable elements are exactly
those that differ by a dimensionless factor, leading to the following
corollary:

\begin{corollary}
    Dimensions form a commutative group under multiplication. It is the quotient
    group of the multiplicative group of non-zero elements by the group of
    dimensionless non-zero elements.
\end{corollary}

Here, multiplication of dimensions is defined in the obvious way:
$\tilde{\mathcal{F}}_a \times \tilde{\mathcal{F}}_b \equiv \tilde{\mathcal{F}}_{a \times b}$.
\section{Coherent unit systems and algebras of physical quantities}\label{sec:quantities}

Theorems \ref{dimensionless-field} and \ref{value-unit} establish that
elements of a partially additive field behave very similar to physical
quantities. Dimensionless elements form a field, such as $\mathbb{R}$ or
$\mathbb{C}$. Any element can be uniquely represented as a product of a
``number'' (a dimensionless element) and a unit. Basic arithmetic can be done
in an obvious way provided the unit system forms a group under multiplication
(i. e., the if-clauses in Theorem \ref{value-unit} are satisfied). This
suggests the following definitions:

\begin{definition}
  A unit system is called coherent iff it forms a group under
  multiplication.
\end{definition}

\begin{definition}
  A partially additive field $\tilde{\mathcal{F}}$ is called an algebra of
  physical quantities iff it admits a coherent unit system. Its elements are
  called physical quantities.
\end{definition}

The existence of a coherent unit system may seem a less ``natural'' axiom
than those of a partially additive field, and one may wonder whether it can be
derived from other, more ``natural'' axioms.

One possibility would be to postulate some kind of ``indistinguishability of
units''. For example, we could demand that any two non-zero dimensionful
elements $a, b \in \tilde{\mathcal{F}}$ of the same dimension be related by
some automorphism of $\tilde{\mathcal{F}}$. If we also forbid dimensionful
roots of dimensionless quantities, the existence of a coherent unit system
could be established (see Appendix \ref{sec:coherent} for details).

However, these axioms seem too restrictive for practical purposes. If the
field of dimensionless quantities, ${\tilde{\mathcal{F}}_1} $, is isomorphic
to the field of real numbers, $\mathbb{R}$, then the element $a \times a$ (for
some non-zero dimensionful $a \in \tilde{\mathcal{F}}$) has a square root, but
$(- 1) \times a \times a$ does not, so they are distinguishable.

Another option is to postulate that ${\tilde{\mathcal{F}}_1}  =\mathbb{R}$ or
$\mathbb{C}$ or any other field one wishes to work with, and to forbid
dimensionful roots of dimensionless quantities. Then, if the multiplicative group of
$\tilde{\mathcal{F}}_1$ is cotorsion, such as $\mathbb{R}\backslash \{ 0 \}$
or $\mathbb{C}\backslash \{ 0 \}$,\footnote{Because divisible groups,
finite groups, and their direct products are cotorsion~\cite{encyclopaedia, Fuchs}.} the
existence of a coherent unit system again could be established.\footnote{In
mathematical terms, this means that any extension of a cotorsion group by a
torsion-free group splits~\cite{encyclopaedia, Fuchs}. The torsion-free condition is necessary:
a partially additive field $\mathbb{R} \cup i\mathbb{R}$ consisting of real
and purely imaginary numbers, viewed as having different dimensions, does not
admit a coherent unit system.} A drawback of this approach is that the particular choice of
the field $\tilde{\mathcal{F}}_1$ is artificial.

The philosophical question of what exactly should be the field of
dimensionless quantities does not have an immediately obvious answer. A lot of
quantities that are traditionally assumed real can sometimes be complex in
modern physics. Some examples are imaginary time in finite-temperature field
theory {\cite{imaginarytime,vanLeeuwen}}, complex mass of unstable particles
{\cite{complexmass1,complexmass2,complexmass3}}, complex voltages and currents
arising as amplitudes of harmonic signals {\cite{complexvoltage}}. On the other hand, real numbers
are often sufficient.

For these reasons, we propose not to introduce any axioms restricting the
possible choices for the field of dimensionless quantities, except for
postulating that the underlying multiplicative group is cotorsion.

Yet another option is to postulate that dimensions form a free abelian group
{\cite{Raposo2018}}. In physical terms, every dimension must be uniquely
representable as a finite product of integer powers of base dimensions. Then
it is trivial to construct a coherent unit system by assigning a unit to
each base dimension.

However, allowing only integer exponents seems too restrictive. Fractional
powers of base dimensions can occur in a variety of contexts, including CGS
units for electromagnetic quantities {\cite{deBoer1995review}}, SI unit of
noise-equivalent power (W/$\sqrt{\tmop{Hz}}$), critical
phenomena~\cite{critical_phenomena,table_of_critical_exponents},
fractals~\cite{fractals}, and fractional calculus {\cite{fractional_calculus}}.
In the latter three cases, one might even encounter power laws with
transcendental exponents~\cite{continuous_exponents}. A yet more extreme
example is dimensional regularization in quantum field theory, which may
involve complex exponents {\cite{complex_exponents1,complex_exponents2}}.

In this section, we have discussed various extra axioms that would ensure the
existence of a coherent unit system. Some of them place restrictions on the
field of dimensionless quantities, others on the group of dimensions. Each of
them may have its advantages and disadvantages, so we leave the choice of a
particular axiom as a subject of debate.
\section{Discussion and conclusion}

To summarize, we have shown how the concepts of physical dimensions and
physical units arise naturally from slightly modified axioms of a field if
addition is allowed to be a partial operation. We have also shown how a couple
of extra axioms ensure the existence of a coherent unit system, that allows
one to do arithmetic with physical quantities in the usual way.

Our axiomatic system has an elegant property: quantities, values, units, and
dimensions are all united in a {\tmem{single}} (partial) algebraic structure.
Units are just non-zero quantities, numerical values are dimensionless
quantities, and dimensions are equivalence classes of mutually summable
quantities. This is in contrast to previous works, which usually built the
algebra of quantities upon a commutative group of dimensions or upon the field of
real numbers.

Dimensions could also be identified with zero quantities, because there is a
one-to-one correspondence between them (0 m = 0 cm = zero length). Then
dimensions, units, and values would all be just special kinds of quantities.
However, this would require special axioms defining multiplicative inverses
of zero quantities ($a \times a^{-1} = 0_1$ instead of $a \times a^{-1} = 1$).
We did not adopt this approach because defining special arithmetic for zero
quantities and calling them ``dimensions'' is almost the same as defining
dimensions and quantities separately, which has been done many times before.

As established by Theorems \ref{dimensionless-field} and \ref{value-unit}, our
axioms define basically the same algebraic structure as previous works, just
somewhat generalized by allowing arbitrary fields in place of $\mathbb{R}$ and
an arbitrary (not necessarily finite) number of base dimensions. For this
reason, we did not discuss topics that have already been discussed elsewhere,
such as change of units~\cite{Quade1961,Parkinson1964,Kurth1965,Whitney1968,Krantz1971,Szekeres1978,Carlson1979,Carinena1988,Kitano2013,Domotor2017},
change of base dimensions~\cite{Quade1961,Whitney1968,Szekeres1978,Carlson1979,Carinena1988,Kitano2013,Raposo2018},
or Buckingham's $\Pi$ theorem~\cite{Drobot1953,Quade1961,Parkinson1964,Kurth1965,Whitney1968,Krantz1971,Szekeres1978,Carinena1988,Aragon2004,Kitano2013,Jonsson2014}.

The extra generality allows one to explore unconventional choices for the
field of dimensionless quantities, besides $\mathbb{R}$ and $\mathbb{C}$.
Some interesting options include computable numbers, rational functions, or
meromorphic functions. Thus, partially additive fields can describe not only
scalar quantities, but also scalar fields.

Partially additive fields can be further extended and generalized in various ways:
\begin{itemize}
  \item By introducing an order relation into the axioms in the spirit of
  measurement theory {\cite{Krantz1971}}.
  
  \item By adding axioms of exponentiation that will define non-integer powers
  of quantities.
  
  \item By including non-scalar quantities (such as vectors) {\cite{Hart1995}}
  and quantities belonging to interval scales (such as temperature)~\cite{Stevens}.
\end{itemize}

We consider one of the possible generalizations in Appendix \ref{sec:fieldoids}.
There, we introduce the notion of a {\tmem{fieldoid}} (by analogy with groupoids),
which generalizes a partially additive field by making both addition and
multiplication partial operations. Then we prove that summability implies
multipliability, making a fieldoid essentially a union of partially additive
fields (PAFs), each consisting of mutually multipliable elements. These PAFs
are completely disjoint in the sense that no arithmetic operations can be
performed on quantities from different PAFs.

This offers a possible explanation as to why we can always multiply any two
quantities in standard quantity calculus: non-multipliable quantities can
never appear in the same equation and, therefore, do not belong to a single
physical theory.\footnote{Of course, this is speculation based on a specific
axiomatization. This reasoning will fail in algebraic structures that include
non-scalar and interval-scale quantities, which are not always multipliable.}

On a philosophical note, partially additive fields can be viewed as evidence in
support of Dirac's epistemological views, who believed that physical concepts
can emerge from ``pretty mathematics''~\cite{prettymath}.

As a final remark, our results illustrate that partial algebraic structures are
elegant and powerful tools for describing the physical world.\footnote{Even
a field is a partial algebraic structure, because the multiplicative inverse of
zero is undefined.}
We hope this will stimulate further research in this direction.
\newpage
\begin{acknowledgments}
This paper is written in GNU TeXmacs~\cite{TeXmacs}, a convenient tool for
mathematical typesetting.

The work was supported by the grant \#23-71-01123 of the Russian Science
Foundation, https://rscf.ru/en/project/23-71-01123/.
\end{acknowledgments}


\begin{thebibliography}{53}%
\makeatletter
\providecommand \@ifxundefined [1]{%
 \@ifx{#1\undefined}
}%
\providecommand \@ifnum [1]{%
 \ifnum #1\expandafter \@firstoftwo
 \else \expandafter \@secondoftwo
 \fi
}%
\providecommand \@ifx [1]{%
 \ifx #1\expandafter \@firstoftwo
 \else \expandafter \@secondoftwo
 \fi
}%
\providecommand \natexlab [1]{#1}%
\providecommand \enquote  [1]{``#1''}%
\providecommand \bibnamefont  [1]{#1}%
\providecommand \bibfnamefont [1]{#1}%
\providecommand \citenamefont [1]{#1}%
\providecommand \href@noop [0]{\@secondoftwo}%
\providecommand \href [0]{\begingroup \@sanitize@url \@href}%
\providecommand \@href[1]{\@@startlink{#1}\@@href}%
\providecommand \@@href[1]{\endgroup#1\@@endlink}%
\providecommand \@sanitize@url [0]{\catcode `\\12\catcode `\$12\catcode `\&12\catcode `\#12\catcode `\^12\catcode `\_12\catcode `\%12\relax}%
\providecommand \@@startlink[1]{}%
\providecommand \@@endlink[0]{}%
\providecommand \url  [0]{\begingroup\@sanitize@url \@url }%
\providecommand \@url [1]{\endgroup\@href {#1}{\urlprefix }}%
\providecommand \urlprefix  [0]{URL }%
\providecommand \Eprint [0]{\href }%
\providecommand \doibase [0]{https://doi.org/}%
\providecommand \selectlanguage [0]{\@gobble}%
\providecommand \bibinfo  [0]{\@secondoftwo}%
\providecommand \bibfield  [0]{\@secondoftwo}%
\providecommand \translation [1]{[#1]}%
\providecommand \BibitemOpen [0]{}%
\providecommand \bibitemStop [0]{}%
\providecommand \bibitemNoStop [0]{.\EOS\space}%
\providecommand \EOS [0]{\spacefactor3000\relax}%
\providecommand \BibitemShut  [1]{\csname bibitem#1\endcsname}%
\let\auto@bib@innerbib\@empty
\bibitem [{\citenamefont {Maxwell}(1873)}]{Maxwell}%
  \BibitemOpen
  \bibfield  {author} {\bibinfo {author} {\bibfnamefont {J.~C.}\ \bibnamefont {Maxwell}},\ }\href {https://en.wikisource.org/wiki/A_Treatise_on_Electricity_and_Magnetism/Volume_1/Preliminary} {\emph {\bibinfo {title} {A treatise on electricity and magnetism}}},\ Vol.~\bibinfo {volume} {1}\ (\bibinfo  {publisher} {Clarendon Press},\ \bibinfo {year} {1873})\BibitemShut {NoStop}%
\bibitem [{\citenamefont {Bridgman}(1922)}]{Bridgman}%
  \BibitemOpen
  \bibfield  {author} {\bibinfo {author} {\bibfnamefont {P.~W.}\ \bibnamefont {Bridgman}},\ }\href {https://books.google.com/books?id=pIlCAAAAIAAJ} {\emph {\bibinfo {title} {Dimensional analysis}}}\ (\bibinfo  {publisher} {Yale University Press},\ \bibinfo {year} {1922})\BibitemShut {NoStop}%
\bibitem [{\citenamefont {{de}~Boer}(1995)}]{deBoer1995review}%
  \BibitemOpen
  \bibfield  {author} {\bibinfo {author} {\bibfnamefont {J.}~\bibnamefont {{de}~Boer}},\ }\bibfield  {title} {\enquote {\bibinfo {title} {On the history of quantity calculus and the international system},}\ }\href {https://doi.org/10.1088/0026-1394/31/6/001} {\bibfield  {journal} {\bibinfo  {journal} {Metrologia}\ }\textbf {\bibinfo {volume} {31}},\ \bibinfo {pages} {405} (\bibinfo {year} {1995})}\BibitemShut {NoStop}%
\bibitem [{\citenamefont {Fleischmann}(1951)}]{Fleischmann1951}%
  \BibitemOpen
  \bibfield  {author} {\bibinfo {author} {\bibfnamefont {R.}~\bibnamefont {Fleischmann}},\ }\bibfield  {title} {\enquote {\bibinfo {title} {Die {S}truktur des physikalischen {B}egriffssystems},}\ }\href {https://doi.org/10.1007/BF01379590} {\bibfield  {journal} {\bibinfo  {journal} {Zeitschrift f{\"u}r Physik}\ }\textbf {\bibinfo {volume} {129}},\ \bibinfo {pages} {377--400} (\bibinfo {year} {1951})}\BibitemShut {NoStop}%
\bibitem [{\citenamefont {Drobot}(1953)}]{Drobot1953}%
  \BibitemOpen
  \bibfield  {author} {\bibinfo {author} {\bibfnamefont {S.}~\bibnamefont {Drobot}},\ }\bibfield  {title} {\enquote {\bibinfo {title} {On the foundations of dimensional analysis},}\ }\href {https://doi.org/10.4064/sm-14-1-84-99} {\bibfield  {journal} {\bibinfo  {journal} {Studia Mathematica}\ }\textbf {\bibinfo {volume} {14}},\ \bibinfo {pages} {84--99} (\bibinfo {year} {1953})}\BibitemShut {NoStop}%
\bibitem [{\citenamefont {Quade}(1961)}]{Quade1961}%
  \BibitemOpen
  \bibfield  {author} {\bibinfo {author} {\bibfnamefont {W.}~\bibnamefont {Quade}},\ }\bibfield  {title} {\enquote {\bibinfo {title} {{\"U}ber die algebraische {S}truktur des {G}r{\"o}{\ss}enkalk{\"u}ls der {P}hysik},}\ }\href {https://leopard.tu-braunschweig.de/servlets/MCRFileNodeServlet/dbbs_derivate_00027518/Quade_Groessenkalkuel.pdf} {\bibfield  {journal} {\bibinfo  {journal} {Abhandlungen der Braunschweigischen Wissenschaftlichen Gesellschaft}\ }\textbf {\bibinfo {volume} {13}},\ \bibinfo {pages} {24--65} (\bibinfo {year} {1961})}\BibitemShut {NoStop}%
\bibitem [{\citenamefont {Parkinson}(1964)}]{Parkinson1964}%
  \BibitemOpen
  \bibfield  {author} {\bibinfo {author} {\bibfnamefont {M.}~\bibnamefont {Parkinson}},\ }\bibfield  {title} {\enquote {\bibinfo {title} {An axiomatic approach to dimensions in physics},}\ }\href {https://doi.org/10.1119/1.1970173} {\bibfield  {journal} {\bibinfo  {journal} {American Journal of Physics}\ }\textbf {\bibinfo {volume} {32}},\ \bibinfo {pages} {200--205} (\bibinfo {year} {1964})}\BibitemShut {NoStop}%
\bibitem [{\citenamefont {Kurth}(1965)}]{Kurth1965}%
  \BibitemOpen
  \bibfield  {author} {\bibinfo {author} {\bibfnamefont {R.}~\bibnamefont {Kurth}},\ }\bibfield  {title} {\enquote {\bibinfo {title} {A note on dimensional analysis},}\ }\href {https://doi.org/10.1080/00029890.1965.11970652} {\bibfield  {journal} {\bibinfo  {journal} {The American Mathematical Monthly}\ }\textbf {\bibinfo {volume} {72}},\ \bibinfo {pages} {965--969} (\bibinfo {year} {1965})}\BibitemShut {NoStop}%
\bibitem [{\citenamefont {Whitney}(1968)}]{Whitney1968}%
  \BibitemOpen
  \bibfield  {author} {\bibinfo {author} {\bibfnamefont {H.}~\bibnamefont {Whitney}},\ }\bibfield  {title} {\enquote {\bibinfo {title} {The mathematics of physical quantities: {P}art {II}: {Q}uantity structures and dimensional analysis},}\ }\href {https://doi.org/10.1080/00029890.1968.11970972} {\bibfield  {journal} {\bibinfo  {journal} {The American Mathematical Monthly}\ }\textbf {\bibinfo {volume} {75}},\ \bibinfo {pages} {227--256} (\bibinfo {year} {1968})}\BibitemShut {NoStop}%
\bibitem [{\citenamefont {Steiner}(1969)}]{Steiner1969}%
  \BibitemOpen
  \bibfield  {author} {\bibinfo {author} {\bibfnamefont {H.-G.}\ \bibnamefont {Steiner}},\ }\bibfield  {title} {\enquote {\bibinfo {title} {Magnitudes and rational numbers — {A} didactical analysis},}\ }\href {https://doi.org/10.1007/BF00303470} {\bibfield  {journal} {\bibinfo  {journal} {Educational Studies in Mathematics}\ }\textbf {\bibinfo {volume} {2}},\ \bibinfo {pages} {371--392} (\bibinfo {year} {1969})}\BibitemShut {NoStop}%
\bibitem [{\citenamefont {Krantz}\ \emph {et~al.}(1971)\citenamefont {Krantz}, \citenamefont {Luce}, \citenamefont {Suppes},\ and\ \citenamefont {Tversky}}]{Krantz1971}%
  \BibitemOpen
  \bibfield  {author} {\bibinfo {author} {\bibfnamefont {D.}~\bibnamefont {Krantz}}, \bibinfo {author} {\bibfnamefont {D.}~\bibnamefont {Luce}}, \bibinfo {author} {\bibfnamefont {P.}~\bibnamefont {Suppes}},\ and\ \bibinfo {author} {\bibfnamefont {A.}~\bibnamefont {Tversky}},\ }\href {https://books.google.com/books?id=aZ3bmPjCc9QC&pg=PA461} {\emph {\bibinfo {title} {Foundations of measurement, Vol. {I}: {A}dditive and polynomial representations}}}\ (\bibinfo  {publisher} {Academic Press},\ \bibinfo {year} {1971})\BibitemShut {NoStop}%
\bibitem [{\citenamefont {Alten}\ and\ \citenamefont {Quade}(1971)}]{Quade1971}%
  \BibitemOpen
  \bibfield  {author} {\bibinfo {author} {\bibfnamefont {H.-W.}\ \bibnamefont {Alten}}\ and\ \bibinfo {author} {\bibfnamefont {W.}~\bibnamefont {Quade}},\ }\bibfield  {title} {\enquote {\bibinfo {title} {Analysis im {G}r{\"o}{\ss}enkalk{\"u}l der {P}hysik},}\ }\href {https://leopard.tu-braunschweig.de/servlets/MCRFileNodeServlet/dbbs_derivate_00030530/Alten_Quade_Analysis_im_Groessenkalkuel_der_Physik.pdf} {\bibfield  {journal} {\bibinfo  {journal} {Abhandlungen der Braunschweigischen Wissenschaftlichen Gesellschaft}\ }\textbf {\bibinfo {volume} {23}},\ \bibinfo {pages} {311--341} (\bibinfo {year} {1971})}\BibitemShut {NoStop}%
\bibitem [{\citenamefont {Bunge}(1971)}]{Bunge1971}%
  \BibitemOpen
  \bibfield  {author} {\bibinfo {author} {\bibfnamefont {M.}~\bibnamefont {Bunge}},\ }\bibfield  {title} {\enquote {\bibinfo {title} {A mathematical theory of the dimensions and units of physical quantities},}\ }in\ \href {https://doi.org/10.1007/978-3-642-80624-7_1} {\emph {\bibinfo {booktitle} {Problems in the Foundations of Physics}}}\ (\bibinfo  {publisher} {Springer},\ \bibinfo {year} {1971})\ pp.\ \bibinfo {pages} {1--16}\BibitemShut {NoStop}%
\bibitem [{\citenamefont {Page}(1975)}]{Page1975}%
  \BibitemOpen
  \bibfield  {author} {\bibinfo {author} {\bibfnamefont {C.~H.}\ \bibnamefont {Page}},\ }\bibfield  {title} {\enquote {\bibinfo {title} {On the dimension group of classical physics},}\ }\href {https://doi.org/10.6028/JRES.079B.013} {\bibfield  {journal} {\bibinfo  {journal} {J. Res. Natl. Bur. Stand. B}\ }\textbf {\bibinfo {volume} {79}},\ \bibinfo {pages} {127--13} (\bibinfo {year} {1975})}\BibitemShut {NoStop}%
\bibitem [{\citenamefont {Brunk}\ and\ \citenamefont {Myszkowski}(1975)}]{Brunk1975}%
  \BibitemOpen
  \bibfield  {author} {\bibinfo {author} {\bibfnamefont {G.}~\bibnamefont {Brunk}}\ and\ \bibinfo {author} {\bibfnamefont {J.}~\bibnamefont {Myszkowski}},\ }\bibfield  {title} {\enquote {\bibinfo {title} {Zur {A}lgebra physikalischer {G}r{\"o}{\ss}en und {M}a{\ss}einheiten},}\ }\href {https://doi.org/10.1007/BF02560794} {\bibfield  {journal} {\bibinfo  {journal} {Forschung im Ingenieurwesen A}\ }\textbf {\bibinfo {volume} {41}},\ \bibinfo {pages} {154--158} (\bibinfo {year} {1975})}\BibitemShut {NoStop}%
\bibitem [{\citenamefont {Szekeres}(1978)}]{Szekeres1978}%
  \BibitemOpen
  \bibfield  {author} {\bibinfo {author} {\bibfnamefont {P.}~\bibnamefont {Szekeres}},\ }\bibfield  {title} {\enquote {\bibinfo {title} {The mathematical foundations of dimensional analysis and the question of fundamental units},}\ }\href {https://www.academia.edu/54708988/The_mathematical_foundations_of_dimensional_analysis_and_the_question_of_fundamental_units} {\bibfield  {journal} {\bibinfo  {journal} {International Journal of Theoretical Physics}\ }\textbf {\bibinfo {volume} {17}},\ \bibinfo {pages} {957--974} (\bibinfo {year} {1978})}\BibitemShut {NoStop}%
\bibitem [{\citenamefont {Carlson}(1979)}]{Carlson1979}%
  \BibitemOpen
  \bibfield  {author} {\bibinfo {author} {\bibfnamefont {D.~E.}\ \bibnamefont {Carlson}},\ }\bibfield  {title} {\enquote {\bibinfo {title} {A mathematical theory of physical units, dimensions, and measures},}\ }\href {https://doi.org/10.1007/BF00281156} {\bibfield  {journal} {\bibinfo  {journal} {Archive for Rational Mechanics and Analysis}\ }\textbf {\bibinfo {volume} {70}},\ \bibinfo {pages} {289--305} (\bibinfo {year} {1979})}\BibitemShut {NoStop}%
\bibitem [{\citenamefont {Rychlewski}(1980)}]{Rychlewski1980}%
  \BibitemOpen
  \bibfield  {author} {\bibinfo {author} {\bibfnamefont {J.}~\bibnamefont {Rychlewski}},\ }\bibfield  {title} {\enquote {\bibinfo {title} {Mathematical foundations of the theory of dimensions and similarity},}\ }\href {https://vjs.ac.vn/index.php/vjmech/article/view/9908/pdf} {\bibfield  {journal} {\bibinfo  {journal} {Vietnam Journal of Mechanics}\ }\textbf {\bibinfo {volume} {2}},\ \bibinfo {pages} {1--8} (\bibinfo {year} {1980})}\BibitemShut {NoStop}%
\bibitem [{\citenamefont {Houard}(1981)}]{Houard1981}%
  \BibitemOpen
  \bibfield  {author} {\bibinfo {author} {\bibfnamefont {J.-C.}\ \bibnamefont {Houard}},\ }\bibfield  {title} {\enquote {\bibinfo {title} {Sur la description intrins{\`e}que des grandeurs dimensionnelles},}\ }\href {http://www.numdam.org/item/AIHPA_1981__35_3_225_0.pdf\#page=11} {\bibfield  {journal} {\bibinfo  {journal} {Annales de l'institut Henri Poincar{\'e}. Section A, Physique Th{\'e}orique}\ }\textbf {\bibinfo {volume} {35}},\ \bibinfo {pages} {225--252} (\bibinfo {year} {1981})}\BibitemShut {NoStop}%
\bibitem [{\citenamefont {Cari{\~n}ena}\ and\ \citenamefont {Santander}(1988)}]{Carinena1988}%
  \BibitemOpen
  \bibfield  {author} {\bibinfo {author} {\bibfnamefont {J.~F.}\ \bibnamefont {Cari{\~n}ena}}\ and\ \bibinfo {author} {\bibfnamefont {M.}~\bibnamefont {Santander}},\ }\bibfield  {title} {\enquote {\bibinfo {title} {Dimensional analysis},}\ }in\ \href {https://doi.org/10.1016/S0065-2539(08)60559-4} {\emph {\bibinfo {booktitle} {Advances in Electronics and Electron Physics}}},\ Vol.~\bibinfo {volume} {72}\ (\bibinfo  {publisher} {Elsevier},\ \bibinfo {year} {1988})\ pp.\ \bibinfo {pages} {181--258}\BibitemShut {NoStop}%
\bibitem [{\citenamefont {Hart}(1995)}]{Hart1995}%
  \BibitemOpen
  \bibfield  {author} {\bibinfo {author} {\bibfnamefont {G.~W.}\ \bibnamefont {Hart}},\ }\href {https://books.google.com/books?id=zoWuORTkBwwC&pg=PA30} {\emph {\bibinfo {title} {Multidimensional analysis: {A}lgebras and systems for science and engineering}}}\ (\bibinfo  {publisher} {Springer-Verlag},\ \bibinfo {year} {1995})\BibitemShut {NoStop}%
\bibitem [{\citenamefont {Aragon}(2004)}]{Aragon2004}%
  \BibitemOpen
  \bibfield  {author} {\bibinfo {author} {\bibfnamefont {S.}~\bibnamefont {Aragon}},\ }\bibfield  {title} {\enquote {\bibinfo {title} {The algebraic structure of physical quantities},}\ }\href {https://www.academia.edu/49269455/The_Algebraic_Structure_of_Physical_Quantities} {\bibfield  {journal} {\bibinfo  {journal} {Journal of mathematical chemistry}\ }\textbf {\bibinfo {volume} {36}},\ \bibinfo {pages} {55--74} (\bibinfo {year} {2004})}\BibitemShut {NoStop}%
\bibitem [{\citenamefont {Pescetti}(2009)}]{Pescetti2009}%
  \BibitemOpen
  \bibfield  {author} {\bibinfo {author} {\bibfnamefont {D.}~\bibnamefont {Pescetti}},\ }\bibfield  {title} {\enquote {\bibinfo {title} {Dimensional analysis and qualitative methods in problem solving: {II}},}\ }\href {https://doi.org/10.1088/0143-0807/30/5/016} {\bibfield  {journal} {\bibinfo  {journal} {European journal of physics}\ }\textbf {\bibinfo {volume} {30}},\ \bibinfo {pages} {1093} (\bibinfo {year} {2009})}\BibitemShut {NoStop}%
\bibitem [{\citenamefont {Sharlow}(2009)}]{Sharlow2009}%
  \BibitemOpen
  \bibfield  {author} {\bibinfo {author} {\bibfnamefont {M.}~\bibnamefont {Sharlow}},\ }\href {https://philpapers.org/archive/SHAGTA.pdf\#page=5} {\enquote {\bibinfo {title} {Generalizing the algebra of physical quantities},}\ }\bibinfo {howpublished} {preprint} (\bibinfo {year} {2009})\BibitemShut {NoStop}%
\bibitem [{\citenamefont {Jany{\v{s}}ka}, \citenamefont {Modugno},\ and\ \citenamefont {Vitolo}(2010)}]{Janyska2010}%
  \BibitemOpen
  \bibfield  {author} {\bibinfo {author} {\bibfnamefont {J.}~\bibnamefont {Jany{\v{s}}ka}}, \bibinfo {author} {\bibfnamefont {M.}~\bibnamefont {Modugno}},\ and\ \bibinfo {author} {\bibfnamefont {R.}~\bibnamefont {Vitolo}},\ }\bibfield  {title} {\enquote {\bibinfo {title} {An algebraic approach to physical scales},}\ }\href {https://inspirehep.net/files/95008b7b74d08273494b066da013d18b} {\bibfield  {journal} {\bibinfo  {journal} {Acta applicandae mathematicae}\ }\textbf {\bibinfo {volume} {110}},\ \bibinfo {pages} {1249--1276} (\bibinfo {year} {2010})}\BibitemShut {NoStop}%
\bibitem [{\citenamefont {Tao}(2012)}]{Tao2012}%
  \BibitemOpen
  \bibfield  {author} {\bibinfo {author} {\bibfnamefont {T.}~\bibnamefont {Tao}},\ }\href {https://terrytao.wordpress.com/2012/12/29/a-mathematical-formalisation-of-dimensional-analysis} {\enquote {\bibinfo {title} {A mathematical formalisation of dimensional analysis},}\ }\bibinfo {howpublished} {blog post} (\bibinfo {year} {2012})\BibitemShut {NoStop}%
\bibitem [{\citenamefont {Kitano}(2013)}]{Kitano2013}%
  \BibitemOpen
  \bibfield  {author} {\bibinfo {author} {\bibfnamefont {M.}~\bibnamefont {Kitano}},\ }\bibfield  {title} {\enquote {\bibinfo {title} {Mathematical structure of unit systems},}\ }\href {https://arxiv.org/pdf/1305.1291} {\bibfield  {journal} {\bibinfo  {journal} {Journal of Mathematical Physics}\ }\textbf {\bibinfo {volume} {54}} (\bibinfo {year} {2013})}\BibitemShut {NoStop}%
\bibitem [{\citenamefont {Domotor}(2017)}]{Domotor2017}%
  \BibitemOpen
  \bibfield  {author} {\bibinfo {author} {\bibfnamefont {Z.}~\bibnamefont {Domotor}},\ }\bibfield  {title} {\enquote {\bibinfo {title} {Torsor theory of physical quantities and their measurement},}\ }\href {https://www.measurement.sk/2017/msr-2017-0019.pdf} {\bibfield  {journal} {\bibinfo  {journal} {Measurement Science Review}\ }\textbf {\bibinfo {volume} {17}},\ \bibinfo {pages} {152--177} (\bibinfo {year} {2017})}\BibitemShut {NoStop}%
\bibitem [{\citenamefont {Raposo}(2018)}]{Raposo2018}%
  \BibitemOpen
  \bibfield  {author} {\bibinfo {author} {\bibfnamefont {{\'A}.~P.}\ \bibnamefont {Raposo}},\ }\bibfield  {title} {\enquote {\bibinfo {title} {The algebraic structure of quantity calculus},}\ }\href {https://www.measurement.sk/2018/msr-2018-0021.pdf\#page=3} {\bibfield  {journal} {\bibinfo  {journal} {Measurement Science Review}\ }\textbf {\bibinfo {volume} {18}},\ \bibinfo {pages} {147--157} (\bibinfo {year} {2018})}\BibitemShut {NoStop}%
\bibitem [{\citenamefont {Zapata-Carratala}(2022)}]{ZapataCarratala2022}%
  \BibitemOpen
  \bibfield  {author} {\bibinfo {author} {\bibfnamefont {C.}~\bibnamefont {Zapata-Carratala}},\ }\bibfield  {title} {\enquote {\bibinfo {title} {Dimensioned algebra: {M}athematics with physical quantities},}\ }\href {https://arxiv.org/pdf/2108.08703} {\bibfield  {journal} {\bibinfo  {journal} {La Matematica}\ }\textbf {\bibinfo {volume} {1}},\ \bibinfo {pages} {849--885} (\bibinfo {year} {2022})}\BibitemShut {NoStop}%
\bibitem [{\citenamefont {Jonsson}(2014)}]{Jonsson2014}%
  \BibitemOpen
  \bibfield  {author} {\bibinfo {author} {\bibfnamefont {D.}~\bibnamefont {Jonsson}},\ }\href {https://arxiv.org/pdf/1408.5024\#page=3} {\enquote {\bibinfo {title} {Quantities, dimensions and dimensional analysis},}\ }\bibinfo {howpublished} {preprint} (\bibinfo {year} {2014}),\ \Eprint {https://arxiv.org/abs/1408.5024} {arXiv:1408.5024} \BibitemShut {NoStop}%
\bibitem [{\citenamefont {Jonsson}(2023)}]{Jonsson2023}%
  \BibitemOpen
  \bibfield  {author} {\bibinfo {author} {\bibfnamefont {D.}~\bibnamefont {Jonsson}},\ }\bibfield  {title} {\enquote {\bibinfo {title} {Scalable monoids and quantity calculus},}\ }\href {https://doi.org/10.1007/s00233-023-10371-0} {\bibfield  {journal} {\bibinfo  {journal} {Semigroup Forum}\ }\textbf {\bibinfo {volume} {107}},\ \bibinfo {pages} {158--187} (\bibinfo {year} {2023})}\BibitemShut {NoStop}%
\bibitem [{\citenamefont {Semple}\ and\ \citenamefont {Whittle}(1996)}]{partialfields}%
  \BibitemOpen
  \bibfield  {author} {\bibinfo {author} {\bibfnamefont {C.}~\bibnamefont {Semple}}\ and\ \bibinfo {author} {\bibfnamefont {G.}~\bibnamefont {Whittle}},\ }\bibfield  {title} {\enquote {\bibinfo {title} {Partial fields and matroid representation},}\ }\href {https://doi.org/10.1006/aama.1996.0010} {\bibfield  {journal} {\bibinfo  {journal} {Advances in Applied Mathematics}\ }\textbf {\bibinfo {volume} {17}},\ \bibinfo {pages} {184--208} (\bibinfo {year} {1996})}\BibitemShut {NoStop}%
\bibitem [{\citenamefont {Hazewinkel}(1997)}]{encyclopaedia}%
  \BibitemOpen
  \bibfield  {author} {\bibinfo {author} {\bibfnamefont {M.}~\bibnamefont {Hazewinkel}},\ }\href {https://encyclopediaofmath.org/wiki/Cotorsion_group} {\emph {\bibinfo {title} {Encyclopaedia of Mathematics: {S}upplement Volume {I}}}}\ (\bibinfo  {publisher} {Springer},\ \bibinfo {year} {1997})\BibitemShut {NoStop}%
\bibitem [{\citenamefont {Fuchs}(2015)}]{Fuchs}%
  \BibitemOpen
  \bibfield  {author} {\bibinfo {author} {\bibfnamefont {L.}~\bibnamefont {Fuchs}},\ }\href {https://books.google.com/books?id=2KMvCwAAQBAJ&pg=PA136} {\emph {\bibinfo {title} {Abelian groups}}}\ (\bibinfo  {publisher} {Springer},\ \bibinfo {year} {2015})\BibitemShut {NoStop}%
\bibitem [{\citenamefont {Landsman}\ and\ \citenamefont {{van}~Weert}(1987)}]{imaginarytime}%
  \BibitemOpen
  \bibfield  {author} {\bibinfo {author} {\bibfnamefont {N.~P.}\ \bibnamefont {Landsman}}\ and\ \bibinfo {author} {\bibfnamefont {C.~G.}\ \bibnamefont {{van}~Weert}},\ }\bibfield  {title} {\enquote {\bibinfo {title} {Real- and imaginary-time field theory at finite temperature and density},}\ }\href {https://doi.org/10.1016/0370-1573(87)90121-9} {\bibfield  {journal} {\bibinfo  {journal} {Physics Reports}\ }\textbf {\bibinfo {volume} {145}},\ \bibinfo {pages} {141--249} (\bibinfo {year} {1987})}\BibitemShut {NoStop}%
\bibitem [{\citenamefont {Stefanucci}\ and\ \citenamefont {{van}~Leeuwen}(2013)}]{vanLeeuwen}%
  \BibitemOpen
  \bibfield  {author} {\bibinfo {author} {\bibfnamefont {G.}~\bibnamefont {Stefanucci}}\ and\ \bibinfo {author} {\bibfnamefont {R.}~\bibnamefont {{van}~Leeuwen}},\ }\href {https://books.google.com/books?id=6GsrjPFXLDYC&pg=PA104} {\emph {\bibinfo {title} {Nonequilibrium many-body theory of quantum systems: {A} modern introduction}}}\ (\bibinfo  {publisher} {Cambridge University Press},\ \bibinfo {year} {2013})\BibitemShut {NoStop}%
\bibitem [{\citenamefont {Matthews}\ and\ \citenamefont {Salam}(1958)}]{complexmass1}%
  \BibitemOpen
  \bibfield  {author} {\bibinfo {author} {\bibfnamefont {P.}~\bibnamefont {Matthews}}\ and\ \bibinfo {author} {\bibfnamefont {A.}~\bibnamefont {Salam}},\ }\bibfield  {title} {\enquote {\bibinfo {title} {Relativistic field theory of unstable particles},}\ }\href {https://doi.org/10.1103/PhysRev.112.283} {\bibfield  {journal} {\bibinfo  {journal} {Physical Review}\ }\textbf {\bibinfo {volume} {112}},\ \bibinfo {pages} {283} (\bibinfo {year} {1958})}\BibitemShut {NoStop}%
\bibitem [{\citenamefont {Zwanziger}(1963)}]{complexmass2}%
  \BibitemOpen
  \bibfield  {author} {\bibinfo {author} {\bibfnamefont {D.}~\bibnamefont {Zwanziger}},\ }\bibfield  {title} {\enquote {\bibinfo {title} {Unstable particles in {S}-matrix theory},}\ }\href {https://doi.org/10.1103/PhysRev.131.888} {\bibfield  {journal} {\bibinfo  {journal} {Physical Review}\ }\textbf {\bibinfo {volume} {131}},\ \bibinfo {pages} {888} (\bibinfo {year} {1963})}\BibitemShut {NoStop}%
\bibitem [{\citenamefont {Denner}\ and\ \citenamefont {Dittmaier}(2006)}]{complexmass3}%
  \BibitemOpen
  \bibfield  {author} {\bibinfo {author} {\bibfnamefont {A.}~\bibnamefont {Denner}}\ and\ \bibinfo {author} {\bibfnamefont {S.}~\bibnamefont {Dittmaier}},\ }\bibfield  {title} {\enquote {\bibinfo {title} {The complex-mass scheme for perturbative calculations with unstable particles},}\ }\href {https://arxiv.org/pdf/hep-ph/0605312} {\bibfield  {journal} {\bibinfo  {journal} {Nuclear Physics B - Proceedings Supplements}\ }\textbf {\bibinfo {volume} {160}},\ \bibinfo {pages} {22--26} (\bibinfo {year} {2006})},\ \bibinfo {note} {proceedings of the 8th DESY Workshop on Elementary Particle Theory}\BibitemShut {NoStop}%
\bibitem [{\citenamefont {Steinmetz}(1893)}]{complexvoltage}%
  \BibitemOpen
  \bibfield  {author} {\bibinfo {author} {\bibfnamefont {C.~P.}\ \bibnamefont {Steinmetz}},\ }\bibfield  {title} {\enquote {\bibinfo {title} {Complex quantities and their use in electrical engineering},}\ }in\ \href {https://books.google.com/books?id=8p8EAAAAYAAJ&pg=PA40} {\emph {\bibinfo {booktitle} {Proceedings of the International Electric Congress, {h}eld in Chicago, 1893}}}\ (\bibinfo {year} {1893})\ pp.\ \bibinfo {pages} {33--75}\BibitemShut {NoStop}%
\bibitem [{\citenamefont {Nishimori}\ and\ \citenamefont {Ortiz}(2010)}]{critical_phenomena}%
  \BibitemOpen
  \bibfield  {author} {\bibinfo {author} {\bibfnamefont {H.}~\bibnamefont {Nishimori}}\ and\ \bibinfo {author} {\bibfnamefont {G.}~\bibnamefont {Ortiz}},\ }\href {https://books.google.com/books?id=F5syAwAAQBAJ} {\emph {\bibinfo {title} {Elements of phase transitions and critical phenomena}}}\ (\bibinfo  {publisher} {Oxford University Press},\ \bibinfo {year} {2010})\BibitemShut {NoStop}%
\bibitem [{\citenamefont {Pelissetto}\ and\ \citenamefont {Vicari}(2002)}]{table_of_critical_exponents}%
  \BibitemOpen
  \bibfield  {author} {\bibinfo {author} {\bibfnamefont {A.}~\bibnamefont {Pelissetto}}\ and\ \bibinfo {author} {\bibfnamefont {E.}~\bibnamefont {Vicari}},\ }\bibfield  {title} {\enquote {\bibinfo {title} {Critical phenomena and renormalization-group theory},}\ }\href {https://arxiv.org/pdf/cond-mat/0012164\#page=43} {\bibfield  {journal} {\bibinfo  {journal} {Physics Reports}\ }\textbf {\bibinfo {volume} {368}},\ \bibinfo {pages} {549--727} (\bibinfo {year} {2002})}\BibitemShut {NoStop}%
\bibitem [{\citenamefont {Bunde}\ and\ \citenamefont {Havlin}(2013)}]{fractals}%
  \BibitemOpen
  \bibfield  {author} {\bibinfo {author} {\bibfnamefont {A.}~\bibnamefont {Bunde}}\ and\ \bibinfo {author} {\bibfnamefont {S.}~\bibnamefont {Havlin}},\ }\href {https://books.google.com/books?id=dh7rCAAAQBAJ&pg=PA232} {\emph {\bibinfo {title} {Fractals in science}}}\ (\bibinfo  {publisher} {Springer-Verlag},\ \bibinfo {year} {2013})\BibitemShut {NoStop}%
\bibitem [{\citenamefont {West}(2014)}]{fractional_calculus}%
  \BibitemOpen
  \bibfield  {author} {\bibinfo {author} {\bibfnamefont {B.~J.}\ \bibnamefont {West}},\ }\bibfield  {title} {\enquote {\bibinfo {title} {Colloquium: {F}ractional calculus view of complexity: {A} tutorial},}\ }\href {https://doi.org/10.1103/RevModPhys.86.1169} {\bibfield  {journal} {\bibinfo  {journal} {Reviews of Modern Physics}\ }\textbf {\bibinfo {volume} {86}},\ \bibinfo {pages} {1169--1186} (\bibinfo {year} {2014})}\BibitemShut {NoStop}%
\bibitem [{\citenamefont {Andrade}\ and\ \citenamefont {Herrmann}(2013)}]{continuous_exponents}%
  \BibitemOpen
  \bibfield  {author} {\bibinfo {author} {\bibfnamefont {R.~F.~S.}\ \bibnamefont {Andrade}}\ and\ \bibinfo {author} {\bibfnamefont {H.~J.}\ \bibnamefont {Herrmann}},\ }\bibfield  {title} {\enquote {\bibinfo {title} {Percolation model with continuously varying exponents},}\ }\href {https://arxiv.org/pdf/1507.07614} {\bibfield  {journal} {\bibinfo  {journal} {Physical Review E—Statistical, Nonlinear, and Soft Matter Physics}\ }\textbf {\bibinfo {volume} {88}},\ \bibinfo {pages} {042122} (\bibinfo {year} {2013})}\BibitemShut {NoStop}%
\bibitem [{\citenamefont {Ashmore}(1972)}]{complex_exponents1}%
  \BibitemOpen
  \bibfield  {author} {\bibinfo {author} {\bibfnamefont {J.~F.}\ \bibnamefont {Ashmore}},\ }\bibfield  {title} {\enquote {\bibinfo {title} {A method of gauge-invariant regularization},}\ }\href {https://doi.org/10.1007/BF02824407} {\bibfield  {journal} {\bibinfo  {journal} {Lettere al Nuovo Cimento (1971-1985)}\ }\textbf {\bibinfo {volume} {4}},\ \bibinfo {pages} {289--290} (\bibinfo {year} {1972})}\BibitemShut {NoStop}%
\bibitem [{\citenamefont {Capper}\ and\ \citenamefont {Leibbrandt}(1974)}]{complex_exponents2}%
  \BibitemOpen
  \bibfield  {author} {\bibinfo {author} {\bibfnamefont {D.}~\bibnamefont {Capper}}\ and\ \bibinfo {author} {\bibfnamefont {G.}~\bibnamefont {Leibbrandt}},\ }\bibfield  {title} {\enquote {\bibinfo {title} {Dimensional regularization for zero-mass particles in quantum field theory},}\ }\href {https://inis.iaea.org/records/bvgvs-szg16/files/4067037.pdf\#page=9} {\bibfield  {journal} {\bibinfo  {journal} {Journal of Mathematical Physics}\ }\textbf {\bibinfo {volume} {15}},\ \bibinfo {pages} {82--85} (\bibinfo {year} {1974})}\BibitemShut {NoStop}%
\bibitem [{\citenamefont {Stevens}(1946)}]{Stevens}%
  \BibitemOpen
  \bibfield  {author} {\bibinfo {author} {\bibfnamefont {S.~S.}\ \bibnamefont {Stevens}},\ }\bibfield  {title} {\enquote {\bibinfo {title} {On the theory of scales of measurement},}\ }\href {https://dl.icdst.org/pdfs/files3/38396a3ee53de9478a57af4605f0e1c1.pdf} {\bibfield  {journal} {\bibinfo  {journal} {Science}\ }\textbf {\bibinfo {volume} {103}},\ \bibinfo {pages} {677--680} (\bibinfo {year} {1946})}\BibitemShut {NoStop}%
\bibitem [{\citenamefont {Dirac}(1982)}]{prettymath}%
  \BibitemOpen
  \bibfield  {author} {\bibinfo {author} {\bibfnamefont {P.~A.}\ \bibnamefont {Dirac}},\ }\bibfield  {title} {\enquote {\bibinfo {title} {Pretty mathematics},}\ }\href {https://doi.org/10.1007/BF02650229} {\bibfield  {journal} {\bibinfo  {journal} {International journal of theoretical physics}\ }\textbf {\bibinfo {volume} {21}},\ \bibinfo {pages} {603--605} (\bibinfo {year} {1982})}\BibitemShut {NoStop}%
\bibitem [{\citenamefont {{GNU TeXmacs}}()}]{TeXmacs}%
  \BibitemOpen
  \bibfield  {author} {\bibinfo {author} {\bibnamefont {{GNU TeXmacs}}},\ }\href@noop {} {}\bibinfo {howpublished} {\href{https://www.texmacs.org/tmweb/home/videos.en.html}{https://www.texmacs.org/}}\BibitemShut {NoStop}%
\bibitem [{\citenamefont {Weinstein}(1996)}]{groupoids1}%
  \BibitemOpen
  \bibfield  {author} {\bibinfo {author} {\bibfnamefont {A.}~\bibnamefont {Weinstein}},\ }\bibfield  {title} {\enquote {\bibinfo {title} {Groupoids: {U}nifying internal and external symmetry. {A} tour through some examples},}\ }\href {https://www.ams.org/notices/199607/weinstein.pdf\#page=3} {\bibfield  {journal} {\bibinfo  {journal} {Notices of the AMS}\ }\textbf {\bibinfo {volume} {43}},\ \bibinfo {pages} {744--752} (\bibinfo {year} {1996})}\BibitemShut {NoStop}%
\bibitem [{\citenamefont {Ivan}(2002)}]{groupoids2}%
  \BibitemOpen
  \bibfield  {author} {\bibinfo {author} {\bibfnamefont {G.}~\bibnamefont {Ivan}},\ }\bibfield  {title} {\enquote {\bibinfo {title} {Algebraic constructions of {B}randt groupoids},}\ }in\ \href {https://px-pict.com/articles/brandt/07_ivan.pdf\#page=2} {\emph {\bibinfo {booktitle} {Proceedings of the Algebra Symposium, Babeș-Bolyai University, Cluj}}}\ (\bibinfo {year} {2002})\ pp.\ \bibinfo {pages} {69--90}\BibitemShut {NoStop}%
\end{thebibliography}
%

\appendix
\section{Fieldoids}\label{sec:fieldoids}

In this section we show that if we relax the axioms of a partially additive
field by making {\tmem{both}} addition and multiplication partial operations,
we get essentially a union of disjoint partially additive fields. We will
call such an algebraic structure a {\tmem{fieldoid}}, by analogy with
groupoids {\cite{groupoids1,groupoids2}}.

\newpage

\begin{definition}
  Let $\mathcal{Q}_{\mathfrak{u}}$ be a union of two disjoint non-empty sets
  $\mathcal{Q}$ and $\mathcal{U}$, with $\mathcal{U}$ containing a single
  element, $\mathfrak{u}$. Let $+ : \mathcal{Q}_{\mathfrak{u}} \otimes
  \mathcal{Q}_{\mathfrak{u}} \rightarrow \mathcal{Q}_{\mathfrak{u}}$, $\times
  : \mathcal{Q}_{\mathfrak{u}} \otimes \mathcal{Q}_{\mathfrak{u}} \rightarrow
  \mathcal{Q}_{\mathfrak{u}}$ be two binary operations on
  $\mathcal{Q}_{\mathfrak{u}}$.
  
  We will use notations $0_a$ and $1_a$ for the additive and multiplicative
  identities corresponding to some $a \in \mathcal{Q}$, and $\mathcal{Z}=
  \left\{ a \in \mathcal{Q}| \exists b \in \mathcal{Q} \quad a = 0_b \right\}$
  for the set of all zero elements of $\mathcal{Q}$.
  
  $\tilde{\mathcal{F}} = (\mathcal{Q}, \mathfrak{u}, +, \times)$ is called a
  fieldoid iff the following axioms are satisfied:
  \begin{eq*}
    \forall a & \in \mathcal{Q}_{\mathfrak{u}} & \mathfrak{u}+ a & = \mathfrak{u}\\
    \forall a & \in \mathcal{Q}_{\mathfrak{u}} & \mathfrak{u} \times a & = \mathfrak{u}\\
    \forall a, b & \in \mathcal{Q}_{\mathfrak{u}} & a + b & = b + a\\
    \forall a, b & \in \mathcal{Q}_{\mathfrak{u}} & a \times b & = b \times a\\
    \forall a, b, c & \in \mathcal{Q}_{\mathfrak{u}} & (a + b) + c & = a + (b + c)\\
    \forall a, b, c & \in \mathcal{Q}_{\mathfrak{u}} & (a \times b) \times c & = a \times (b \times c)\\
    \forall a, b, c & \in \mathcal{Q}_{\mathfrak{u}} & a \times (b + c) & = (a \times b) + (a \times c)\\
    \forall a \in \mathcal{Q} \quad \exists !0_a & \in \mathcal{Q} & a + 0_a & = a\\
    \forall a \in \mathcal{Q}\backslash\mathcal{Z} \quad \exists !1_a & \in \mathcal{Q} & a \times 1_a & = a\\
    \forall a \in \mathcal{Q} \quad \exists (- a) & \in \mathcal{Q} & a + (-a) & = 0_a\\
    \forall a \in \mathcal{Q}\backslash\mathcal{Z} \quad \exists a^{- 1} & \in \mathcal{Q} & a \times a^{- 1} & = 1_a\\
    \forall a \in \mathcal{Z} \quad \exists b & \in \mathcal{Q}\backslash\mathcal{Z} & a & = 0_b
  \end{eq*}
\end{definition}

\begin{lemma}
  $\forall a \in \mathcal{Q} \quad a \times a \neq \mathfrak{u}$ (every
  element is squareable)\label{squareability}.
\end{lemma}

\begin{proof}
  If $a \in \mathcal{Q}\backslash\mathcal{Z}$, then $a = a \times 1_a = a
  \times a \times a^{- 1}$. Therefore, $a \times a \neq \mathfrak{u}$.
  
  If $a \in \mathcal{Z}$, choose $b \in \mathcal{Q}\backslash\mathcal{Z}$ such
  that $a = 0_b$. We have just proven that $b \times b \neq \mathfrak{u}$. But
  $b \times b = (b + a) \times (b + a) = a \times a + \tmop{other}
  \tmop{terms}$, which implies $a \times a \neq \mathfrak{u}$.
\end{proof}

\begin{theorem}
  If two elements of a fieldoid are summable, they are also
  multipliable.\label{summability-multipliability}
\end{theorem}

\begin{proof}
  Assume $a + b \neq \mathfrak{u}$. Then $(a + b) \times (a + b) \neq
  \mathfrak{u}$ by Lemma \ref{squareability}. Since $(a + b) \times (a + b) =
  a \times a + a \times b + b \times a + b \times b$, $a \times b \neq
  \mathfrak{u}$.
\end{proof}

\begin{lemma}
  $\forall a, b \in \mathcal{Q}\backslash\mathcal{Z} \quad a \times b \nin
  \mathcal{Z}$ (no zero(s) divisors).\label{no-divisors-of-zeros-fieldoid}
\end{lemma}

\begin{proof}
  Assume $a \times b = 0_c$ for some $c \in \mathcal{Q}$. Then $c = c + a
  \times b = c + a \times b + a \times b$, so $a \times b = a \times b + a
  \times b$ because $0_c$ must be unique. Multiplying by $b^{- 1}$, we get $a
  \times 1_b = a \times 1_b + a \times 1_b$, so $a \times 1_b$ is its own
  zero. ($a \times b = a \times 1_a \times b = a \times 1_b \times b$ implies
  $a \times 1_b$ and $1_a \times b$ are defined.)
  
  $a \times b + a \times b = a \times b + a \times 1_b \times b = (a + a
  \times 1_b) \times b$, so $a$ is summable with $a \times 1_b$ and they have
  identical zeros (Lemma \ref{identical-zeros} still holds for fieldoids).
  Therefore, $0_a = 0_{a \times 1_b} = a \times 1_b$.
  
  Now we have $a = a + 0_a = a \times 1_a + a \times 1_b = a \times (1_a +
  1_b)$. Because $1_a$ must be unique, $1_a + 1_b = 1_a$. Multiplying by $b$,
  we get $1_a \times b = 1_a \times b + b$. Because $1_a \times b \neq
  \mathfrak{u}$, $b = 0_{1_a \times b} \in \mathcal{Z} $, a contradiction.
\end{proof}

\newpage

\begin{lemma}
  $\forall a \in \mathcal{Q}\backslash\mathcal{Z} \quad 0_a \times 1_a = 0_a$.
\end{lemma}

\begin{proof}
  $a = a \times 1_a = (a + 0_a) \times 1_a = a \times 1_a + 0_a \times 1_a = a
  + 0_a \times 1_a$. Since $0_a$ must be unique, $0_a \times 1_a = 0_a$.
\end{proof}

\begin{lemma}
  $\forall a, b \in \mathcal{Q}\backslash\mathcal{Z} \quad 0_a = 0_b
  \Rightarrow 1_a = 1_b$.
\end{lemma}

\begin{proof}
  $0_a = 0_b$ implies $a$ and $b$ are summable (by Lemma
  \ref{identical-zeros}) and, therefore, multipliable (by Theorem
  \ref{summability-multipliability}). By Lemma \ref{no-divisors-of-zeros-fieldoid}, $a
  \times b$ is non-zero, so $a \times b = a \times 1_a \times b = a \times b
  \times 1_b$ implies $1_a = 1_b = 1_{a \times b}$.
\end{proof}

\begin{corollary}
  It is possible to define multiplicative identities for zero elements as $1_a
  = 1_b$ if $a = 0_b$ and $b \in \mathcal{Q}\backslash\mathcal{Z}$. This
  definition does not depend on the choice of $b$, but $1_a$ is not
  necessarily the only multiplicative identity for $a \in \mathcal{Z}$. From
  now on, we will use this definition when we need a multiplicative identity
  for a zero element.
\end{corollary}

\begin{lemma}
  $a \times b \neq \mathfrak{u} \Leftrightarrow 1_a =
  1_b$.\label{identical-unities}
\end{lemma}

\begin{proof}
  For $a, b \in \mathcal{Q}\backslash\mathcal{Z}$ the proof is similar to that
  of Lemma \ref{identical-zeros}. (Lemma \ref{no-divisors-of-zeros-fieldoid} ensures $a
  \times b \nin \mathcal{Z}$.)
  
  If $a \in \mathcal{Z}$, $b \in \mathcal{Q}\backslash\mathcal{Z}$, we can
  choose $c \in \mathcal{Q}\backslash\mathcal{Z}$ such that $a = 0_c$. The
  products $a \times b$ and $c \times b$ are either both defined or both
  undefined, because $a \times b = 0_c \times b = (c + (- c)) \times b = c
  \times b + (- c) \times b$ and $c \times b = (c + 0_c) \times b = c \times b
  + a \times b$. Therefore, $a \times b \neq \mathfrak{u} \Leftrightarrow c
  \times b \neq \mathfrak{u} \Leftrightarrow 1_c = 1_b \Leftrightarrow 1_a =
  1_b$.
  
  The case when both $a$ and $b$ are zero can be treated similarly.
\end{proof}

\begin{corollary}
  In fieldoids multipliability is transitive.
\end{corollary}

We will use notation $\tilde{\mathcal{F}}_a^{\times} = \{ b \in \mathcal{Q}|b
\times a \neq \mathfrak{u} \}$ for the set of all elements of
$\tilde{\mathcal{F}}$ multipliable with $a \in \mathcal{Q}$.

\begin{lemma}
  Let $a, b \in \tilde{\mathcal{F}}_c^{\times}$ be two elements of a fieldoid
  $\tilde{\mathcal{F}}$ multipliable by some third element $c$. Then $a \times
  b, 0_a, 1_a, - a \in \tilde{\mathcal{F}}_c^{\times}$. If $a \nin
  \mathcal{Z}$, then $a^{- 1} \in \tilde{\mathcal{F}}_c^{\times}$. If $a + b
  \neq \mathfrak{u}$, then $a + b \in
  \tilde{\mathcal{F}}_c^{\times}$.\label{multipliable-subset-is-closed}
\end{lemma}

\begin{proof}
  By transitivity of multipliability, $\tilde{\mathcal{F}}_c^{\times} =
  \tilde{\mathcal{F}}_a^{\times}$ and $a \times b$ is defined.
  
  The statements $a + b, a \times b, 0_a, - a \in
  \tilde{\mathcal{F}}_a^{\times}$ follow from the squareability lemma and the
  following identities:
  \begin{itemize}
    \item $(a + b) \times (a + b) = (a + b) \times a + (a + b) \times b$;
    
    \item $(a \times b) \times (a \times b) = ((a \times b) \times a) \times
    b$;
    
    \item $0_a \times 0_a = 0_a \times a + 0_a \times (- a) = a \times a + (-
    a) \times a + 0_a \times (- a)$.
  \end{itemize}
  $1_a$ and $a^{- 1}$ are multipliable by $a$ by definition.
\end{proof}

\newpage

\begin{theorem}
  A subset of mutually multipliable elements of a fieldoid forms a partially
  additive field. Any two such distinct subsets
  $\tilde{\mathcal{F}}_a^{\times}$, $\tilde{\mathcal{F}}_b^{\times}$ are
  completely disjoint in the sense that both
  $\tilde{\mathcal{F}}_a^{\times} \cup \{ \mathfrak{u} \}$ and
  $\tilde{\mathcal{F}}_b^{\times} \cup \{ \mathfrak{u} \}$ are closed under
  all operations considered in the axioms, and operations involving elements
  from both $\tilde{\mathcal{F}}_a^{\times}$ and
  $\tilde{\mathcal{F}}_b^{\times}$ are undefined (i. e., $\forall c \in
  \tilde{\mathcal{F}}_a^{\times}, d \in \tilde{\mathcal{F}}_b^{\times}$ $c + d
  = c \times d =\mathfrak{u}$).
\end{theorem}

\begin{proof}
  By Lemma \ref{identical-unities}, there exists a single multiplicative
  identity for all elements of such a subset. Lemma
  \ref{multipliable-subset-is-closed} establishes that this subset is closed
  under all operations considered in the axioms (whenever they are defined).
  Together with the axioms of a fieldoid, this implies that this subset
  satisfies the axioms of a partially additive field.
  
  Elements from two such subsets are by definition not mutually multipliable
  and, by Theorem \ref{summability-multipliability}, not summable.
\end{proof}
\section{Sufficient conditions for the existence of a coherent unit
system}\label{sec:coherent}

\begin{lemma}
  Let $\tilde{\mathcal{F}} = (\mathcal{Q}, \mathfrak{u}, +, \times)$ be a
  partially additive field and $a, b \in \mathcal{Q}$ two of its elements.
  Then $a \times b \in \mathcal{Z}$ iff $a \in \mathcal{Z}$ or $b \in
  \mathcal{Z}$.\label{no-divisors-of-zeros-PAF}
\end{lemma}

\begin{proof}
  Assume $a \in \mathcal{Q}\backslash\mathcal{Z}$ and $c + a \times b = c$ for
  some $c \in \mathcal{Q}$. Then $a^{- 1} \times c + b = a^{- 1} \times c$.
  Therefore, $b = 0_{a^{- 1} \times c} \in \mathcal{Z}$.
  
  Assume $c + a = c$ for some $c \in \mathcal{Q}$. Then $c \times b = (c + a)
  \times b = c \times b + a \times b$. Therefore, $a \times b = 0_{c \times b}
  \in \mathcal{Z}$.
\end{proof}

\begin{lemma}
  A product of a dimensionless and a dimensionful element is
  dimensionful.\label{dimensionless-times-dimensionful-is-dimensionful}
\end{lemma}

\begin{proof}
  Let $a \in \tilde{\mathcal{F}}_1, b \in \mathcal{Q}\backslash
  \tilde{\mathcal{F}}_1$. Since $a$ is summable with 1, $a \times b + b = (a +
  1) \times b \neq \mathfrak{u}$. Therefore, $a \times b$ is summable with $b$
  and cannot belong to $\tilde{\mathcal{F}}_1$.
\end{proof}

\begin{lemma}
  Non-zero elements of a partially additive field form a commutative group
  under multiplication.\label{non-zero-group}
\end{lemma}

\begin{proof}
  By Lemma \ref{no-divisors-of-zeros-PAF}, $\mathcal{Q}\backslash\mathcal{Z}$ is
  closed under multiplication. Lemma \ref{no-divisors-of-zeros-PAF} and Lemma
  \ref{non-zero-one} immediately imply that $1 \in
  \mathcal{Q}\backslash\mathcal{Z}$ and $\forall a \in
  \mathcal{Q}\backslash\mathcal{Z} \quad a^{- 1} \in
  \mathcal{Q}\backslash\mathcal{Z}$. Commutativity is in the axioms of a
  partially additive field.
\end{proof}

\newpage

\begin{theorem}
  Let $\tilde{\mathcal{F}} = (\mathcal{Q}, \mathfrak{u}, +, \times)$ be a
  partially additive field satisfying the following two conditions:
  \begin{enumerate}
    \item $\forall n \in \mathbb{N} \quad \forall a \in \mathcal{Q}\backslash
    \tilde{\mathcal{F}}_1 \quad a^n \nin \tilde{\mathcal{F}}_1$ (no
    dimensionful roots of dimensionless elements);
    
    \item $\forall n \in \mathbb{N} \quad \forall a \in \mathcal{Q}\backslash
    \tilde{\mathcal{F}}_1 \quad \forall b, b' \in \tilde{\mathcal{F}}_a
    \backslash\mathcal{Z} \quad \left( \exists c \in \mathcal{Q} \quad c^n = b
    \right) \Rightarrow \left( \exists c' \in \mathcal{Q} \quad (c')^n = b'
    \right)$ (any two non-zero dimensionful elements of the same dimension
    cannot be distinguished on the basis that only one of them has an $n$th
    root).
  \end{enumerate}
  Then $\tilde{\mathcal{F}}$ admits a coherent unit system.
\end{theorem}

\begin{proof}
  We will assume that $\tilde{\mathcal{F}}$ contains at least one dimensionful
  element, otherwise $\mathcal{U}= \{ 1 \}$ is a coherent unit system.
  
  The existence of at least one dimensionful element together with the
  non-triviality axiom $\forall a \in \mathcal{Z} \quad \exists b \in
  \mathcal{Q}\backslash\mathcal{Z} \quad a = 0_b$ implies the existence of at
  least one {\tmem{non-zero}} dimensionful element.
  
  Let $a \in \tilde{\mathcal{F}}_1 \backslash\mathcal{Z}$ be any non-zero
  dimensionless element and $b \in (\mathcal{Q}\backslash
  \tilde{\mathcal{F}}_1) \backslash\mathcal{Z}$ any non-zero dimensionful
  element. Then $b^n$ and $a \times b^n$ are non-zero dimensionful elements by
  Lemmas \ref{dimensionless-times-dimensionful-is-dimensionful},
  \ref{non-zero-group} and condition 1. Therefore, condition 2 implies
  $\exists c \in \mathcal{Q} \quad c^n = a \times b^n$. Then $a = (c \times
  b^{- 1})^n$. Therefore, any non-zero dimensionless element has any root,
  i.e. the multiplicative group of $\tilde{\mathcal{F}}_1
  \backslash\mathcal{Z}$ (non-zero dimensionless elements) is divisible.
  
  As a divisible subgroup of $\mathcal{Q}\backslash\mathcal{Z}$,
  $\tilde{\mathcal{F}}_1 \backslash\mathcal{Z}$ is a direct summand in
  $\mathcal{Q}\backslash\mathcal{Z}$ ({\cite{Fuchs}}, p. 136, Theorem 2.6). In
  other words, there exists a subgroup $\mathcal{U}$ of
  $\mathcal{Q}\backslash\mathcal{Z}$ such that every $a \in
  \mathcal{Q}\backslash\mathcal{Z}$ can be (uniquely) written as $b \times c$
  for some $b \in \tilde{\mathcal{F}}_1 \backslash\mathcal{Z}$ and $c \in
  \mathcal{U}$. Such $c$ is summable with $a$, because $a + c = (b + 1) \times
  c \neq \mathfrak{u}$.
  
  This implies every distinct $\tilde{\mathcal{F}}_a$ contains an element of
  $\mathcal{U}$, and this element is unique (otherwise the factorization $a =
  b \times c$ with $b$ dimensionless and $c \in \mathcal{U}$ would be
  non-unique).
  
  Therefore, $\mathcal{U}$ is a coherent unit system.
\end{proof}

\end{document}